# Simultaneously generating Brillouin microlaser and second harmonic within a lithium niobate microdisk

XIAOCHAO LUO,[1,2,10] CHUNTAO LI,[3,4,10] JINTIAN LIN,[1,2,*] RENHONG GAO,[4] YIFEI YAO,[5] YINGNUO QIU,[1,2] LEI WANG,[5] AND YA CHENG[1,3,4,6,7,8,9,†]

[1]State Key Laboratory of High Field Laser Physics and CAS Center for Excellence in Ultra-Intense Laser Science, Shanghai Institute of Optics and Fine Mechanics (SIOM), Chinese Academy of Sciences (CAS), Shanghai 201800, China
[2]Center of Materials Science and Optoelectronics Engineering, University of Chinese Academy of Sciences, Beijing 100049, China
[3]State Key Laboratory of Precision Spectroscopy, East China Normal University, Shanghai 200062, China
[4]The Extreme Optoelectromechanics Laboratory (XXL), School of Physics and Electronic Science, East China Normal University, Shanghai 200241, China
[5]School of Physics, State Key Laboratory of Crystal Materials, Shandong University, Jinan 250100, China
[6]Shanghai Research Center for Quantum Sciences, Shanghai 201315, China
[7]Hefei National Laboratory, Hefei 230088, China
[8]Collaborative Innovation Center of Extreme Optics, Shanxi University, Taiyuan 030006, China
[9]Collaborative Innovation Center of Light Manipulations and Applications, Shandong Normal University, Jinan 250358, China
[10]Xiaochao Luo and Chuntao Li contributed equally to this work.
*E-mail: jintianlin@siom.ac.cn
†E-mail: ya.cheng@siom.ac.cn

**March, 2025**

We report the simultaneous generation of second-harmonic generation (SHG) and Brillouin microlaser in a high-quality thin-film lithium niobate (TFLN) microdisk resonator. The microdisk is fabricated with ultrahigh-Q factor of $4\times10^6$ by photolithography-assisted chemo-mechanical etching, enabling significant cavity-enhancement effect for boosting nonlinear frequency conversion. Under 1559.632 nm pumping, Brillouin microlaser is demonstrated in the microdisk with Stokes Brillouin shift of ~10 GHz, a low threshold of 1.81 mW, and a fundamental linewidth of 254.365 Hz. Meanwhile, efficient SHG is observed at 779.828 nm with an absolute conversion efficiency of 3.8% at pump level of 3.028 mW. The coexistence of these two nonlinear processes is enabled by the simultaneous confinement of the light and acoustic fileds for effect coupling in the microdisk, which enhances both optomechanical and second-order nonlinear interactions. This research provides new possibilities for integrated multi-frequency laser sources and multifunctional nonlinear photonic devices. © 2025 Optical Society of America

Nonlinear optical effects play a crucial role in integrated photonics, particularly in efficient frequency conversion, optical signal processing, and quantum light sources. Among various nonlinear optical effects, second-harmonic generation (SHG) and stimulated Brillouin scattering (SBS) have been extensively studied due to their unique capabilities. SHG, a classic second-order nonlinear process, is widely used for efficient frequency doubling [1-5], enabling the generation of coherent light at wavelengths that are otherwise inaccessible with conventional lasers. On the other hand, SBS, which arises from the coherent interaction between photons and acoustic phonons, has become a cornerstone for narrow-linewidth lasers, optical frequency comb generation, and low phase-noise microwave signal synthesis [6-12]. Despite their individual importance, existing research has primarily focused on achieving SHG or SBS in isolation, and the simultaneous implementation of these two phenomena on a single photonic device remains unexplored. SHG typically requires a material with a large second-order nonlinear susceptibility and precise phase-matching condition, while SBS relies on strong photoelastic effect for optomechanical coupling and the overlap of optical and acoustic modes. These distinct requirements have made it challenging to achieve both phenomena simultaneously in a single photonic platform.

Here, we bridge this gap by leveraging the unique dual advantages of the thin-film lithium niobate on insulator (LNOI) platform—its strong second-order nonlinearity and pronounced photoelastic effects [13-24]. The simultaneous generation of SHG and Stokes Brillouin laser (SBL) is demonstrated in a suspended LNOI microdisk resonator by dispersion engineering, so as to fulfil the requirements of phase match and energy conservation. When pumped at 1559.632 nm, the microdisk resonator generates a backward-propagating SBL with a fundamental linewidth of 254.365 Hz and a remarkably low pump threshold of 1.81 mW. Simultaneously, efficient SHG is achieved with an output power of 0.1146 mW and an absolutely conversion efficiency of ∼ 3.8 %. This breakthrough not only demonstrates the multiple nonlinear potential of lithium niobate microcavities, but also provides new possibilities for integrating multi-frequency laser sources and multifunctional nonlinear photonic devices.

The on-chip LNOI microdisk was fabricated on a Z-cut lithium niobate thin film using femtosecond laser photolithographic-assisted chemo-mechanical etching (PLACE) technique and the details on the fabrication can be found in our previous work [24,25]. The optical microscope image of the fabricated LNOI suspended microdisk supported by a small silicon dioxide pillar to suppress the leakage of the acoustic modes to the substrate [14,15], is depicted in the inset of Fig. 1. The LNOI microdisk has a diameter of ∼ 117 μm and a thickness of ∼590 nm, for simultaneously fulfilling the requirements of phase match and energy conversation for both the SHG and SBL processes [24].

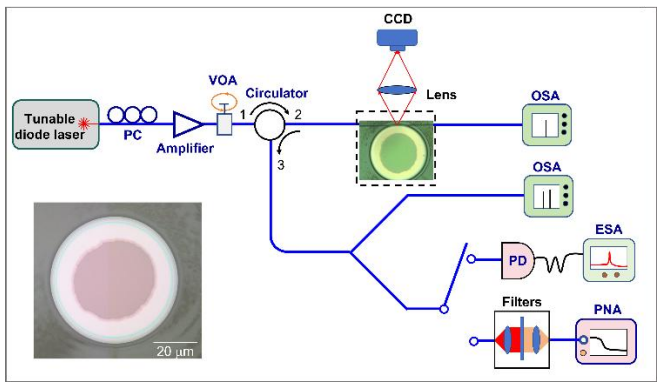

Fig. 1. Schematic of the setup for the formation of SBL and SHG in the microdisk. PC, polarization controller; VOA, variable optical attenuator; CCD, charge coupled device; OSA, optical spectrum analyzer; PD, photodetector; ESA, electrical spectrum analyzer; PNA, phase noise analyzer. Inset: Optical microscope image of the microdisk resonator.

The experimental setup for the formation of SHG and Brillouin microlaser in the suspended LNOI microdisk resonator is illustrated in Fig. 1. A continuous wave (CW) tunable laser (New Focus Inc., Model TLB6728) in the telecom band was used as the pump laser source to excite the nonlinear processes in the microdisk resonator. Before the light from the CW laser was launched in the optical circulator, it propagates through an in-line fiber polarization controller (FPC562, Thorlabs Inc.) and an erbium-doped fiber amplifier (EDFA), followed by a variable optical attenuator (VOA) to modify the polarization, and achieve variable pump power. The pump light was then sent into the microresonator via port 2 of the circulator using an optical tapered fiber with a waist of 2 um. It should be emphasized that the tapered fiber was carefully placed on the edge of the microdisk resonator and was in direct contact with the resonator to couple light in and out of the resonator. The coupling position of the tapered fiber is very crucial for achieving modal phase match for SBS and SHG processes [24,26]. The generated backward Brillouin laser was collected by the circulator, and then detected by an optical spectrum analyzer (AQ6370D, Yokogawa Inc.) for spectral analyses. And a high-speed photodetector connected with a real-time electrical spectrum analyzer (Tektronix RAS5115B) was used to measure the beat note microwave signal by beating the pump light and the SBL. The linewidth of the backward SBL was further measured with a commercially available laser phase noise analyzer (PNA) using tunable optical filters to block the pump light. The generated forward SHG signal was detected by the optical spectral analyzer (OSA).

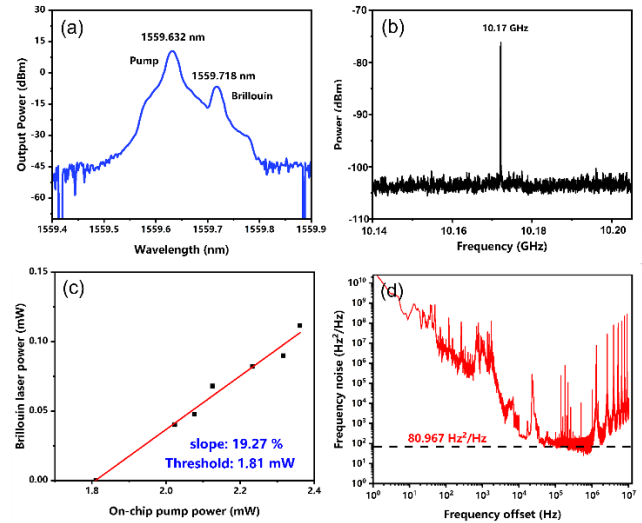

Fig. 2. (a) Optical spectrum of the SBL. (b) Microwave beat signals of the pump wave and the SBL signal. (c) Output power of SBL vs. on-chip pump power. (d) The frequency noise spectrum of the backward SBL.

When the pump light injected into the microdisk resonator was tuned to 1559.632 nm and the pump power exceeded the threshold for exciting the SBL, two distinct nonlinear phenomena were concurrently observed, including a backward-propagating SBL signal and a forward-propagating SHG signal. The backward-propagating SBL signal was detected at 1559.718 nm, as shown in Fig. 2(a). The wavelength interval between the pump light and the SBL signal was approximately 0.08 nm, corresponding to a Stokes Brillouin shift $\Omega_B$ of ∼10 GHz. And this Brillouin shift $\Omega_B$ was further confirmed by the detected radio-frequency (RF) beat note microwave signal with finer resolution by means of optical heterodyne method using ESA. This RF signal centered at 10.17-GHz frequency, as shown in Fig. 2(b). Moreover, the output power of the backward SBL signal varied with different pump powers was also recorded to characterize the threshold behavior of the backward SBL, as plotted in Fig. 2(c). When the pump power exceeds the threshold for SBL, the output power of the backward SBL (black dots) increases linearly with the pump power. Through linear fitting (red line), the threshold of the backward SBL is determined to be 1.81 mW, and the conversion efficiency reaches 19.27%.

Furthermore, the frequency noise of the backward SBL was measured using the correlated self-heterodyne method [12], as depicted in Fig. 2(d). The white-frequency-noise floors $N_{wfn}$ of the backward SBL was quantified as 80.967 Hz$^2$/Hz, revealing short-term linewidths $L_{st}$ of 254.365 Hz ($L_{st} = N_{wfn} \times \pi$) [27]. Such ultra-narrow linewidth is resulted from the narrow Brillouin gain bandwidth and the short lifetime of the Brillouin phonon [7-12,24].

Simultaneously, in the forward-propagating direction, another emission signal was observed at 779.828 nm, as shown in Fig. 3(a). This wavelength is equal to the half of the pump wavelength. As a result, this signal was ascribed to the SHG of the pump light. The on-chip SHG signal reached an output power of 0.1146 mW with an absolute conversion efficiency of 3.8% at an on-chip pump level of 3.028 mW. Notably, no SHG signal in the backward-propagating direction was observed, presumably due to the insufficient power of the backward pump power to excite the SHG process. We also recorded the output power of the SHG signal at different pump powers to characterize the quadratic dependence of the SHG signal on the pump power. The measured SHG conversion efficiency (black dots) increases linearly with the input pump power, as shown in Fig. 3(b). By linear fitting (red line), the normalized SHG conversion efficiency is determined to be 1.3%/mW. Besides the fulfillment of energy conversation rule, modal phase match could be easily realized in the Z-cut microdisk because there are lots of high-Q spatial whispering gallery modes in the second harmonic waveband.

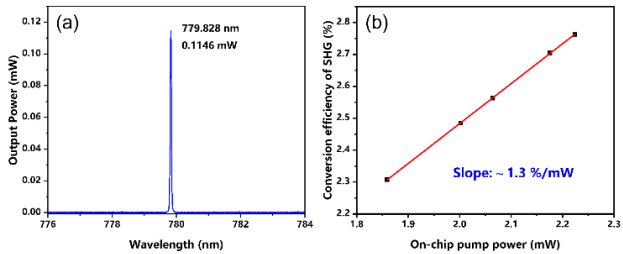

Fig. 3. (a) Spectrum of the SHG. (b) SHG conversion efficiency as a function of the pump power.

To evaluate the Q factor of the microdisk, two continuously tunable lasers were used for exciting the involved modes by scanning the wavelength cross each mode. The pump power injected into the microdisk resonator was set to as low as 5 μW, the transmission spectra of the microdisk resonator were obtained, as shown in Figs. 3(a)-(c). The Lorenz fitting (red curves) shows that the loaded Q factors of the pump mode, Stokes Brillouin mode and second-harmonic (SH) mode were determined to be $4.00\times10^6$, $3.27\times10^6$ and $2.57\times10^6$, respectively. The notably high Q factors guarantee a considerable increase in the intracavity built-up pump power within the small microdisk and boost high conversion efficiencies, which are crucial for low-threshold SBL and efficient SHG.

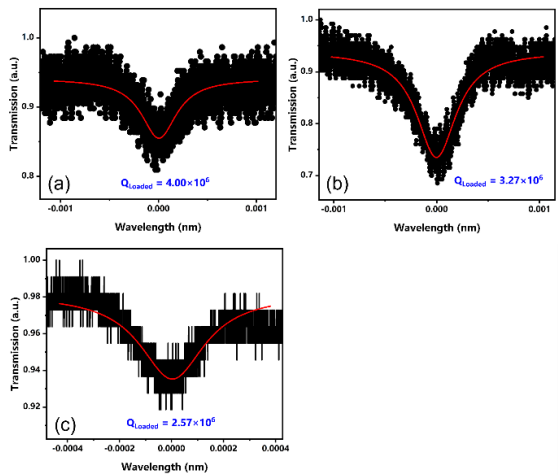

Fig. 4. Transmission spectra of (a) pump mode, (b) Stokes Brillouin mode, and (c) SHG mode, revealing high loaded Q factors through Lorentz fitting (red curves).

In summary, thanks to the strong $\chi^{(2)}$ nonlinearity and pronounced photoelastic effects, both SBL and SHG are observed at the same time in the high-Q LNOI microdisk resonator. The backward SBL is generated with a threshold of only 1.81 mW, which is record-low result demonstrated in LNOI platform so far, accompanied with efficient SHG of a normalized SHG conversion efficiency of 1.3%/mW. By demonstrating the coexistence of Brillouin lasing and SHG in the high-Q LNOI microcavity, we provide a new paradigm for the design of efficient and compact nonlinear photonic integrated systems. This work paves the way for the development of advanced photonic devices that require the integration of diverse functionalities, such as narrow-linewidth lasers, frequency converters, and quantum light sources, all on a single photonic chip.

**Acknowledgments.** We thank Peking university Yangtze delta institute of optoelectronics for providing the laser phase noise measurement system (iFN5000).

**References**

1. I. Breunig, "Three-wave mixing in whispering gallery resonators," Laser Photon. Rev. **10**, 569 (2016).
2. X. Zhang, Q.-T. Cao, Z. Wang, Y.-x. Liu, C.-W. Qiu, L. Yang, Q. Gong, and Y.-F. Xiao, "Symmetry-breaking-induced nonlinear optics at a microcavity surface," Nat. Photon. **13**, 21 (2019).
3. J. Lin, N. Yao, Z. Hao, J. Zhang, W. Mao, M. Wang, W. Chu, R. Wu, Z. Fang, L. Qiao, W. Fang, F. Bo, and Y. Cheng, "Broadband quasi-phaser-matched harmonic generation in an on-chip monocrystalline lithium niobate microdisk resonator," Phys. Rev. Lett. **122**, 173903 (2019).
4. G.-T. Xue, Y.-F. Niu, X. Liu, J-C. Duan, W. Chen, Y. Pan, K, Jia, X. Wang, H.-Y. Liu, Y. Zhang, P. Xu, G. Zhao, X. Cai, Y.-X. Gong, X. Hu, Z. Xie, and S. Zhu, "Ultrabright multiplexed energy-time-entangled photon generation from lithium niobate on insulator chip," Phys. Rev. Appl. **15**, 064059 (2021).
5. Y. Zheng and X. Chen, "Nonlinear wave mixing in lithium niobate thin film," Adv. Phys. X **6**, 1889402 (2021).
6. S. Gundavarapu, G. M. Brodnik, M. Puckett, T. Huffman, D. Bose, R. Behunin, J. Wu, T. Qiu, C. Pinho, N. Chauhan, J. Nohava, P. T. Rakich, K. D. Nelson, M. Salit, and D. J. Blumenthal, "Sub-hertz fundamental linewidth photonic integrated Brillouin laser," Nat. Photonics **13**, 60 (2019).
7. K. Y. Yang, D. Y. Oh, S. H. Lee, Q.-F. Yang, X. Yi, B. Shen, H. Wang, and K. Vahala, "Bridging ultrahigh-Q devices and photonic circuits," Nat. Photonics **12**, 297 (2018).
8. B. J. Eggleton, C. G. Poulton, P. T. Rakich, M. J. Steel, and G. Bahl, "Brillouin integrated photonics," Nat. Photonics **13**, 664 (2019).
9. N. T. Otterstrom, R. O. Behunin, E. A. Kittlaus, Z. Wang, and P. T. Rakich, "A silicon Brillouin laser," Science **360**, 1113 (2018).
10. Y. Bai, M. Zhang, Q. Shi, S. Ding, Y. Qin, Z. Xie, X. Jiang, and M. Xiao, "Brillouin-Kerr soliton frequency combs in an optical microresonator," Phys. Rev. Lett. **126**, 063901 (2021).
11. S. Zhu, B. Xiao, B. Jiang, L. Shi, and X. Zhang, "Tunable Brillouin and Raman microlasers using hybrid microbottle resonators," Nanophotonics **8**, 931 (2019).
12. M. Wang, Z.-G. Hu, C. Lao, Y. Wang, X. Jin, X. Zhou, Y. Lei, Z. Wang, W. Liu, Q.-F. Yang, and B-B. Li, "Taming Brillouin optomechanics using supermode microresonators," Phys. Rev. X **14**, 011056 (2024).
13. Y. Jia, L. Wang, and F. Chen, " Ion-cut lithium niobate on insulator technology: Recent advances and perspectives," Appl. Phys. Rev. **8**, 011307 (2021).

14. D. Zhu, L. Shao, M. Yu, R. Cheng, B. Desiatov, C. J. Xin, Y. Hu, J. Holzgrafe, S. Ghosh, A. Shams-Ansari, E. Puma, N. Sinclair, C. Reimer, M. Zhang, and M. Lončar "Integrated photonics on thin-film lithium niobate," Adv. Opt. Photonics **13**, 242 (2021).
15. G. Chen, N. Li, J. D. Ng, H.-L. Lin, Y. Zhou, Y. H. Fu, L. Y. T. Lee, Y. Yu, A.-Q. Liu, and A. J. Danner, "Advances in lithium niobate photonics development status and perspectives," Adv. Photonics **4**, (2022).
16. J. Lin, F. Bo, Y. Cheng, and J. Xu, "Advances in on-chip photonic devices based on lithium niobate on insulator," Photon. Res. **8**, 1910 (2020).
17. B. Fu, R. Gao, N. Yao, H. Zhang, C. Li, J. Lin, M. Wang, L. Qiao, and Y. Cheng, "Soliton microcomb generation by cavity polygon modes," Opto-Electronic. Adv. **7**, 240061 (2024).
18. C. C. Rodrigues, R. O. Zurita, T. P. M. Alegre, and G. S. Wiederhecker, "Stimulated Brillouin scattering by surface acoustic waves in lithium niobate waveguides," J. Opt. Soc. Am. B **40**, D56–D63 (2023).
19. Y.-H. Yang, J.-Q. Wang, Z.-X. Zhu, X.-B. Xu, Q. Zhang, J. Lu, Y. Zeng, C.-H. Dong, L. Sun, G.-C. Guo, and C.-L. Zou, "Stimulated Brillouin interaction between guided phonons and photons in a lithium niobate waveguide," Sci. China Phys. Mech. Astron. **67**, 214221 (2024).
20. W. Wang, Y. Yu, Y. Li, Z. Bai, G. Wang, K. Li, C. Song, Z. Wang, S. Li, Y. Wang, Z. Lu, Y. Li, T. Liu, X. Yan, Tailorable Brillouin light scattering in a lithium niobate waveguide. Appl. Sci. **11**, 8390 (2021).
21. M. Li, L. Chang, L. Wu, J. Staffa, J. Ling, U. A. Javid, S. Xue, Y. He, R. Lopez-Rios, T. J. Morin, H. Wang, B. Shen, S. Zeng, L. Zhu, K. J. Vahala, J. E. Bowers, and Q. Lin, "Integrated pockels laser," Nat. Commun. **13**, 5344 (2022).
22. J. Lu, J. B. Surya, X. Liu, et al, "Periodically poled thin-film lithium niobate microring resonators with a second-harmonic generation efficiency of 250000%/W," Optica **6**, 1455 (2019).
23. Q. Luo, F. Bo, Y. Kong, G. Zhang, and J. Xu, "Advances in lithium niobate thin-film lasers and amplifiers: a review," Adv. Photon. **3**, 034002 (2023).
24. C. Li, J. Deng, X. Huang, X. Luo, R. Gao, J. Lin, H. Yu, J. Guan, Z. Li, and Y. Cheng, "Sub-kilohertz intrinsic linewidth stimulated Brillouin laser in integrated lithium niobate microresonators," arXiv preprint arXiv:2411.17443 (2024).
25. R. Wu, J. Zhang, N. Yao, W. Fang, L. Qiao, Z. Chai, J. Lin, and Y. Cheng, "Lithium niobate micro-disk resonators of quality factors above $10^7$," Opt. Lett. **43**, 4116 (2018).
26. C. Sun, J. Ni, C. Li, J. Lin, R. Gao, J. Guan, Q. Qiao, Q. Hou, X. Luo, X. Zheng, L. Qiao, M. Wang, and Y. Cheng, "Second harmonic generation with 48% conversion efficiency from cavity polygon modes in a monocrystalline lithium niobate microdisk resonator," Laser Photon. Rev. 2401857 (2025).
27. D. Xu, F. Yang, D. Chen, F. Wei, H. Cai, Z. Fang, and R. Qu, "Laser phase and frequency noise measurement by Michelson interferometer composed of a 3 × 3 optical fiber coupler," Opt. Express **23**, 22386 (2015).